# Sub-atomic movements of a domain wall in the Peierls potential

K.S. Novoselov[1], A.K. Geim[1], S.V. Dubonos[1,3], E.W. Hill[2] and I.V. Grigorieva[1]

[1]Department of Physics & [2]Department of Computer Sciences,
University of Manchester, M13 9PL, Manchester, UK

[3]Institute for Microelectronics Technology, 142432 Chernogolovka, Russia

Movements of individual domain walls in a ferromagnetic garnet were studied with angstrom resolution. The measurements reveal that domain walls can be locked between adjacent crystallographic planes and propagate by distinct steps matching the lattice periodicity. Domain walls are found to be weakly mobile within valleys of the atomic washboard but become unexpectedly flexible on Peierls ridges, where they can be kept in a bi-stable state by ac magnetic field. We describe the latter observation in terms of a single magnetic kink propagating along a domain wall.

The discrete nature of the crystal lattice bears on virtually every material property but it is only when the size of condensed-matter objects - e.g., dislocations [1-3], vortices in superconductors [4-6], domain walls [7,8] - becomes comparable to the lattice period, that the discreteness reveals itself explicitly. The associated phenomena are usually described in terms of the Peierls ("atomic washboard") potential, which was first introduced for the case of dislocations at the dawn of the condensed-matter era [1,2]. Since then, the concept has been invoked in many situations to explain certain features in bulk behavior of various materials but so far eluded direct detection and experimental scrutiny on the microscopic level.

In the particular case of magnetic materials, a domain walls (DW) has to pass through different spin configurations as it moves from one atomic plane to another [7-10]. Fig. 1a shows two principal configurations for a DW in a simple spin lattice, which have the maximum and minimum energy and correspond to the center of the wall lying at and between atomic planes, respectively. This spatial variation of wall's energy is generally referred to as the Peierls potential. Its amplitude depends on the ratio between DW's width $\delta$ and the lattice periodicity $d$ and, realistically, the Peierls potential is only observable for $\delta/d < 10$. For larger $\delta/d$, the potential becomes so small ($\ll 1G$) that pinning on defects should conceal it completely. The vast theoretical and experimental evidence gathered over several decades and based on studies of bulk properties of magnetic rare-earth alloys has confirmed the existence of the Peierls potential, with probably the most definite conclusions drawn from measurements of magnetic viscosity (e.g., see [10-12]).

In the present work, we revisit the Peierls potential by employing a state-of-the-art technique of ballistic Hall micromagnetometry [13] imported from another research area (mesoscopic superconductivity) [14], which allows us to resolve sub-atomic changes in the position of micron-sized segments of DWs and study their inter- and intra- Peierls valley movements. We clearly see that a domain wall can become trapped between crystalline planes, which results in its propagation by clear jumps corresponding to the periodicity of the Peierls potential. DWs are rather rigid at the bottom of Peierls valleys but can also be stabilized in a highly flexible transient state on top of a Peierls ridge.

For our experiments, we have used a 10 μm thick films of yttrium-iron garnet $(YBi)_3(FeGa)_5O_{12}$ (YIG) grown in [111] direction. The films exhibit a saturation magnetization $4\pi M_s$ of ≈200G, exchange energy $A$ of ≈$1.8 \times 10^{-7}$ erg/cm and, below 10K, crystal anisotropy $K$ of ≈$1.4 \times 10^6$ erg/cm$^3$. The above values were found with accuracy ≈10%. The experimental system combines relatively narrow walls (thickness $\delta = \pi(A/K)^{1/2}$ ≈11 nm at helium temperatures) with a large unit cell of size $a$ ≈1.24 nm and provides $\delta/d$ ≈6. Equally important is the high crystal quality of our samples [15-17] manifested in a coercivity of $\ll 0.1$G at room temperature and $< 10$ G at helium temperatures, such that obscuring effects due to pinning on defects are relatively small. The YIG films have perpendicular magnetization and a domain structure shown in Fig. 1b. A sub-mm piece of the film was placed in immediate contact with the surface of a device consisting of several micron-sized Hall sensors made from a two-dimensional electron gas (2DEG) following the microfabrication procedure described in refs [13,14] (Fig. 1b).

Due to in-plane crystal anisotropy, DWs in our YIG films tend to lie in three equivalent planes {110}. This crystallographic alignment is already seen in Fig. 1b at 300K, and it becomes stronger as the anisotropy increases at lower temperatures, where domain walls become straight over distances of many mm. Using alignment marks, we placed our 2DEG sensors inside a chosen area of YIG film with many parallel domains and aligned the sensors parallel to them, i.e. perpendicular to one of <110> axes. The spacing between the garnet and 2DEG was measured to be less than 100 nm [17]. We restricted the reported experiments to temperatures below 30K, mainly because of thermally activated relaxation processes, which led to irreproducible changes in the domain structure and did not allow accurate measurements that require slow sweeps of magnetic field.

In our experiments, submicron Hall probes play a role of highly sensitive position detectors, which provide a spatial resolution of < 1Å with respect to DW movements. When a DW enters the sensitive area of a probe, its response $R_H$ starts changing. A shift $\Delta x$ in DW's position leads to a change in magnetic field $B$ and flux $\Phi$ through the Hall cross (Fig. 1c), which in turn induces a Hall response such that $\Delta R_H = \alpha \Delta \Phi = \beta \Delta x$ [13,18]. The second part of the equation assumes that a DW is straight within the sensitive area of the Hall cross. This is justified for the reported experiments because we could simultaneously measure $R_H$ at different crosses and, at low $T$, found a nearly perfect correlation between movements of DWs detected by neighbouring Hall crosses [17]. This indicates that DWs move as rigid, straight objects so that their large segments (for $T < 10$ K, we have estimated their size to be up to 10 microns) shift as a whole (i.e. without bending). $\alpha$ and $\beta$ are found experimentally [13,17,18]. For brevity, we discuss only experiments where the DWs were aligned parallel with the Hall device, as shown in Fig. 1b, and moved in <110> direction. In this well-defined geometry, changes in the wall position $\Delta x$ can be calculated from changes in $R_H$ directly, without using any fitting parameters.

We used Hall sensors made from a high-mobility 2DEG because of their exceptional sensitivity to flux variations $\delta \Phi$ on a submicron scale. At temperatures $T < 80$ K, such sensors effectively work as fluxmeters and are capable of resolving $\delta \Phi \approx 10^{-4} \phi_0$, where $\phi_0$ is a flux quantum. This technique has previously been used in studies of submicron superconducting [13,14] and ferromagnetic [19-22] particles, and its detailed description can be found therein. In the context of the present work, we have exploited this unique flux sensitivity to achieve an angstrom resolution in the average position of individual DWs. Indeed, if a DW passes the whole width $w$ of a cross, $\Phi$ changes by several $\phi_0$ (for the given value of $M_s$ in our garnets). On the other hand, we can resolve $\delta \Phi \approx 10^{-4} \phi_0$ and this corresponds to a shift of a DW by $\Delta x \approx w(\delta \Phi/\phi_0) \approx 1$Å. Note that many magnetic materials have larger values of $M_s$ and, hence, the magnetometers should then provide even higher spatial resolution ($< 0.1$Å).

To move a DW, we slowly varied external field $H$ applied perpendicular to the garnet film. Figure 2 shows a typical example of changes in local field $B$ detected by a Hall sensor as a DW crosses it from one side to the other. One can see that $B$ changes its sign, which reflects the change in polarity of the domain above the sensor, and zero $B$ corresponds to the state where the wall lies exactly in the middle. The overall shape of the transition curve is in good agreement with a simple theory [17]. Overlaid on this universal behavior, one can see a number of small sample- and sweep- dependent steps, indicating that a DW does not move smoothly but covers micron-long distances in a series of small jumps. Such jumps have previously been studied by many techniques (e.g., [23-25,17]) and are usually referred to as Barkhausen noise. A typical step in Fig. 2 corresponds to a wall moving by 10 to 50 nm. While a DW was located within the Hall cross, we could reverse a field sweep to investigate local coercivity of the wall (left inset in Fig. 2). Such hysteresis loops are usually reproducible for many field cycles, and we attribute them to pinning on individual defects [17,24].

In addition to the above behavior, the high resolution of the micromagnetometry allowed us to discern very small DW jumps (right inset in Fig. 2) which stood out from the "ordinary" ones for two reasons. Firstly, they matched closely the lattice periodicity in the direction of DW travel <110> ($d = 2^{1/2} a \approx 1.75$ nm) and, secondly, they were practically the only jumps observed in the range below $\approx 10$ nm. The use of statistical analysis techniques (standard in e.g. particle physics) shows that - with a confidence level of 94% - the right inset corresponds to an event comprising several steps of equal length (4 single and 3 double steps), where the length of a single step is $d$. To obtain a further proof that such steps indeed reveal jumps between equivalent crystal lattice positions, we carried out complementary experiments described below.

As a DW moves through a crystal, it interacts with a large number of pinning sites and becomes bent and strained in the process. For a strained wall, one can generally expect that it would jump between strong pinning sites without noticing the weaker ones. This is clearly seen by magnetic force microscopy

(at room temperature). To release the strain, we demagnetized the sample by applying an ac magnetic field with an amplitude $h$ gradually decreasing from $\approx$5 G to zero, while a constant field $H$ kept the wall close to the center of the probe. This proved to be a critical improvement: in the demagnetized state, DWs started to propagate via clear quantized jumps matching the lattice periodicity. The distance between the equivalent sites was measured to be 1.6 ±0.2 nm, in agreement with the Peierls potential periodicity $d \approx$1.75 nm. Figure 3 shows an example of such behavior, which leaves no doubt of the presence of a periodic atomic landscape impeding DW movements.

We note that there are two periodic sets of equivalent crystallographic positions for a {110} DW, which are separated by $b$ and $2b$ (where $b = a/2^{1/2}$ is the distance between the nearest basal planes). They require the translation of wall's spin configuration in directions <001> and <110>, respectively, and involve different exchange interactions [9]. Both periods should contribute to the Peierls potential but because of the exponential dependence on $\delta/d$ only the longest periodicity $d = 2a/2^{1/2} \approx$1.75 nm can be expected to remain observable. Our measurements did find this periodicity but it remains unclear why a DW could not avoid the observed Peierls barriers by exploiting "the third dimension" (i.e. moving by twice smaller oblique jumps in <001> rather than by the straight jumps perpendicular to DW's plane). It is also possible that the complex unit cell structure of garnets also plays some role in defining the dominant period.

From Fig. 3, one can estimate that it requires a field of $\approx$1G to move a DW out of the well created by adjacent crystal planes (i.e. intrinsic pinning is several times weaker than pinning on a typical defect in our garnets; see Fig. 2). Further experiments yielded a value of the intrinsic coercive field $H_C \approx$0.7 G at $T$ < 10 K. Theory of the magnetic Peierls potential predicts $H_C$ to be of the order of [9-11]

$$H_C \approx C\,(A/d^2 M_S)\exp(-\pi\delta/d)$$

where constant $C$ is $\approx 10^3$ [9]. Taking into account the exponential dependence of $H_C$ on $\delta$, which is known to $\approx$10% accuracy, the formula yields $H_C$ in the range from 0.1 to 5 G, in agreement with the experiment.

In addition to the detection of the Peierls potential, we have studied its shape, which is predicted to be sinusoidal [7,9]. The latter implies that a DW should remain somewhat mobile within Peierls valleys, i.e. not pinned completely. Such intra-valley movements are expected to be $\approx$1Å and, therefore, could not be resolved in our dc magnetization data (cf. Fig. 3). To gain information about the finest DW movements, we measured local ac susceptibility $\chi$ (ac measurements provide a higher flux sensitivity and hence a higher spatial resolution). To this end, in addition to the dc field $H$ that controls DW's position, we applied an oscillating field $h$ and measured an ac signal generated by oscillatory movements of a DW. Changes in $\chi$ show how the mobility of a DW varies with its position inside a Peierls valley. Using this approach, we confirmed that a domain wall could indeed move near the bottom of a valley, and the detected ac signal corresponded to an average shift of a DW by up to $\approx$0.5Å. In addition to this, however, ac measurements revealed strikingly unusual DW dynamics, in qualitative disagreement with the behavior expected for an object moving in a tilted sinusoidal potential.

One of the most notable features we observed is a large well-reproduced peak in DW mobility shown in Fig. 4. Here, zero $H$ corresponds to a DW position in the middle between two adjacent Peierls valleys, as simultaneously detected in the dc magnetization signal (the latter shows a smeared transition between two DW positions separated by $d$). The peak has abrupt edges (dashed lines), i.e. above a certain value of $H$ the oscillating wall suddenly falls into a neighboring Peierls valley and becomes locked there (i.e. moves by <1Å). When $h$ was switched off for a few seconds at a constant $H$ close to one of the peak's edges, the transient state did not recover on switching the modulation back. This indicates that it requires time of $\approx$1s for the wall to become trapped inside a Peierls valley. With decreasing $h$ from 0.5 to 0.1 G, both the width and the amplitude of the peak shrink linearly (inset in Fig. 4). The observed behavior suggests that ac field stabilizes a DW in what otherwise should be an intrinsically unstable state between two Peierls valleys.

One can interpret the transient state as the center of a DW sitting effectively on a Peierls ridge, kept there by an oscillating magnetic force. This situation closely resembles the so-called reversed pendulum, which can be stabilized in the unstable upside-down position by an oscillating force [26]. This analogy allows us to describe the observed resonance semi-quantitatively but does not provide a microscopic picture. To this end, one can invoke the well-known "kink" model [2,10], where a DW moves between

Peierls valleys via a process where at first only a submicron segment of a DW (jog) moves to the next valley. Spreading the boundary of such jog along the wall eventually leads to the relocation of the whole DW. It is plausible that ac modulation could stabilize the submicron jog so that its boundaries move back and forth inside the sensitive area of a Hall cross without collapsing, until changes in $H$ extend the jog outside of this area, where it eventually becomes pinned. Indeed, the maximum value of $\Delta\chi$ (observed at 30 K and 0.5 G) corresponds to DW movements by $\Delta x \approx 1$nm, i.e. nearly the whole segment of the wall inside the Hall cross swings between adjacent valleys.

The single-jog model provides a sensible description for the behavior in Fig. 4 as well as for the majority of other ac susceptibility results (to be reported elsewhere). However, the origin of the long characteristic times remains puzzling. Moreover, the kink/jog picture may no longer be justifiable for the case of sub-atomic displacements ($\Delta x << \delta$) because, on this scale, the spin configuration of a DW changes (DW should "breathe") and one cannot simply refer to an average shift of a DW as a whole. We believe that the detected transient state can indicate some internal modes excited inside a DW when it is softened and ready to move from one Peierls valley to another.

Further theoretical and experimental work is required to understand the unexpected atomic-scale dynamics of DWs. This physics has been extensively discussed in theory (e.g., solitons on discrete lattices [27-29]) but so far was not accessible in a direct experiment to allow tests for a variety of theoretical models. The reported results show a possibility to study the physics of individual domain walls and solitons within them at a new level of experimental resolution. This should lead not only to refinement of the existing models but also to a greater depth of understanding of fundamental and technologically important phenomena governed by movements of domain walls.

1. R.E. Peierls, *Proc. Phys. Soc.* **52**, 34 (1940).
2. F.R.N. Nabarro, *Proc. Phys. Soc.* **59**, 256 (1947).
3. H.R. Kolar, J.C.H. Spence, H. Alexander, *Phys. Rev. Lett.* **77**, 4031 (1996).
4. B.I. Ivlev, N.B. Kopnin, *Phys. Rev. Lett.* **64**, 1828 (1990).
5. M. Oussena, P.A.J. de Groot, R. Gagnon, L. Taillefer, *Phys. Rev. Lett.* **72**, 3606 (1994).
6. R. Kleiner, F. Steinmeyer, G. Kunkel, P. Müller, *Phys. Rev. Lett.* **68**, 2394 (1992).
7. B. Barbara, *J. Phys*. **34**, 1039 (1973).
8. J.J. van den Broek, H. Zijlstra, *IEEE Trans. Magn.* **7**, 226 (1971).
9. H.R. Hilzinger, H. Kronmüller, *Phys. Stat. Sol.* (b) **54**, 593 (1972).
10. T. Egami, *Phys. Stat. Sol.* (a) **19**, 747 (1973).
11. J.I. Arnaudas, A. del Moral, J.S. Abell, *J. Magn. Magn. Mater.* **61**, 370 (1986).
12. N.V. Musbniskov *et al*, *J. Alloys Comp.* **305**, 188 (2002).
13. A.K. Geim *et al.*, *Appl. Phys. Lett.* **71**, 2379 (1997).
14. A.K. Geim *et al.*, *Nature* **390**, 259 (1997).
15. R.V. Pisarev *et al.*, *J. Phys.* **5**, 8621 (1993).
16. V.V. Pavlov, R.V. Pisarev, A. Kirilyuk, T. Rasing, *Phys. Rev. Lett.* **78**, 2004 (1997).
17. K.S. Novoselov *et al.*, *IEEE Trans. Magn.* **38**, 2583 (2002).
18. F.M. Peeters, X.Q. Li, *Appl. Phys. Lett.* **72**, 572 (1998).
19. A.D. Kent, S. von Molnar, S. Gider, D.D. Awschalom, *J. Appl. Phys.* **76**, 6656 (1994).
20. Y.Q. Li *et al.*, *Appl. Phys. Lett.* **80**, 4644 (2002).
21. T.M. Hengstmann, D. Grundler, C. Heyn, D. Heitmann, *J. Appl. Phys.* **90**, 6542 (2001).
22. D. Schuh *et al*, *IEEE Trans. Magn.* **37**, 2091 (2001).
23. R. Vergne, J.C. Cotillard, J.L. Porteseil, *Rev. Phys. Appl.* **16**, 449 (1981).
24. J. Wunderlich *et al.*, *IEEE Trans. Magn.* **37**, 2104 (2001).
25. D.H. Kim, S.B. Choe, S.C. Shin, *Phys. Rev. Lett.* **90**, 087203 (2003).
26. K. Magnus, *Vibrations.* Blackie &Son, London, 1965.
27. S. Aubry, *J. Phys.* **44**, 147 (1983).
28. O.M. Braun, Y.S. Kivshar, *Phys. Rep.* **306**, 2 (1998).
29. A.R. Bishop, W.F. Lewis, *J. Phys.* C **12**, 3811 (1979).

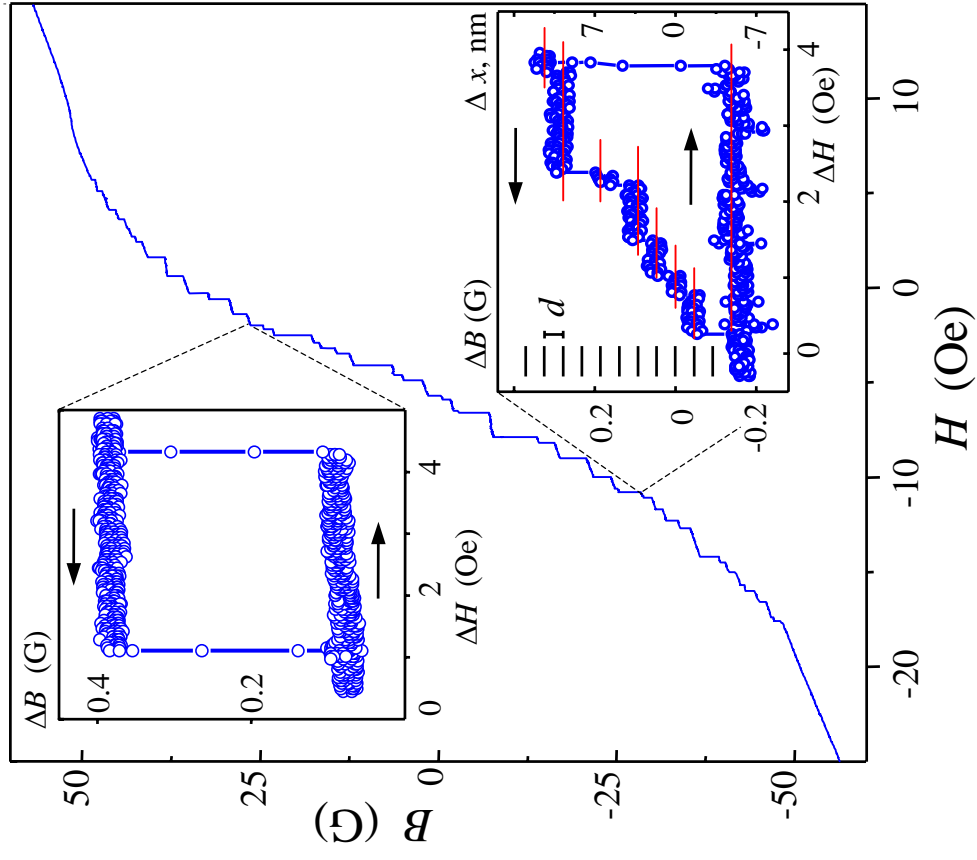

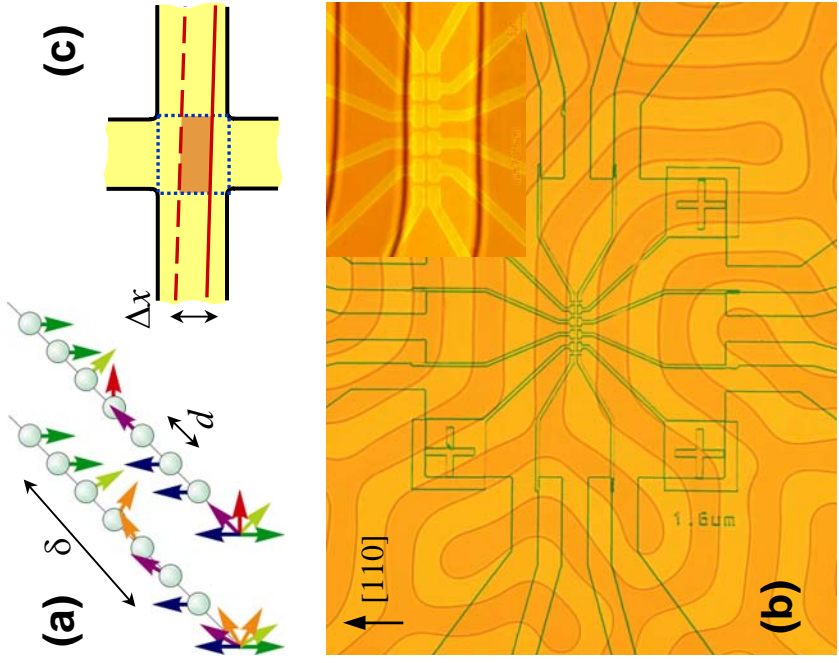

**Figure 1. Experimental structures and devices. a,** Principal spin configurations for a narrow Bloch wall: its centre either coincides with one of the atomic planes (right) or lies between them (left). The fan diagrams show the orientation of individual spins if one looks in the direction perpendicular to DW. **b,** The micrographs show a set of micron-sized Hall probes placed on top of a ferromagnetic garnet. Their image overlays a photograph of the domain structure taken in transmitted polarized light at room temperature. The inset magnifies the central part of the experimental structure. The scale is given by domains' width of ≈14 μm and the size of Hall probes (1.5 μm). By measuring simultaneously the response at different Hall crosses, we ensured that at low temperatures the studied DWs were parallel to the set of sensors as the photo shows. **c,** The drawing illustrates that a shift in the average position of a wall $\Delta x$ induces a change in flux $\Delta\Phi$ inside the sensitive area marked by the dotted lines, which is recorded as a change in Hall resistance [13,18].

**Figure 2. Nanometer movements of domain walls over submicron distances.** The central plot shows a typical Hall response measured by a 1.5 μm Hall cross as a domain wall slowly creeps from one of its sides to the other ($T = 0.5$ K). For convenience, the Hall response is plotted in terms of the average local field $B$ inside the cross, which is calculated by using the measured Hall coefficient. The insets show examples of local hysteresis loops.

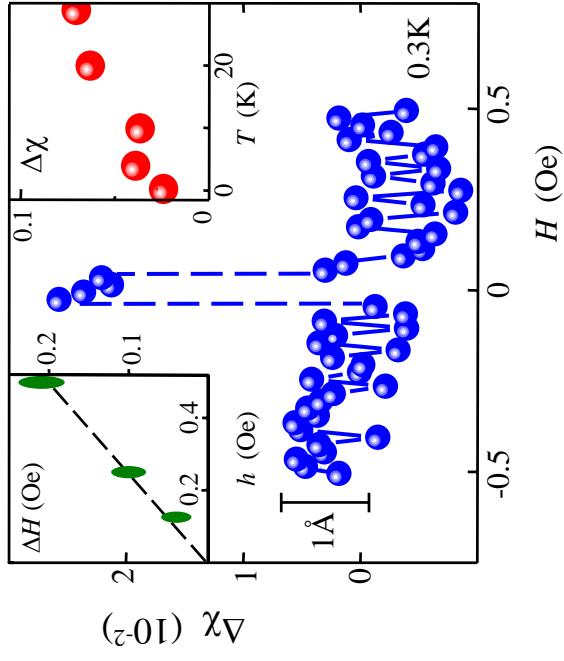

**Figure 4. Domain wall on the Peierls ridge.** The main plot shows changes in local ac susceptibility $\chi$ while dc field $H$ moves the wall from one Peierls valley to the next (ac field has amplitude 0.5G and frequency 8 Hz). We subtracted a constant background due to the Hall response induced directly by the ac field. The slight variation of $\chi$ seen on the curve away from the transient state is not reproducible for different walls and after thermal cycling. The insets show the dependence of the width $\Delta H$ of the transient state on $h$ and the dependence of its amplitude $\Delta \chi$ on temperature.

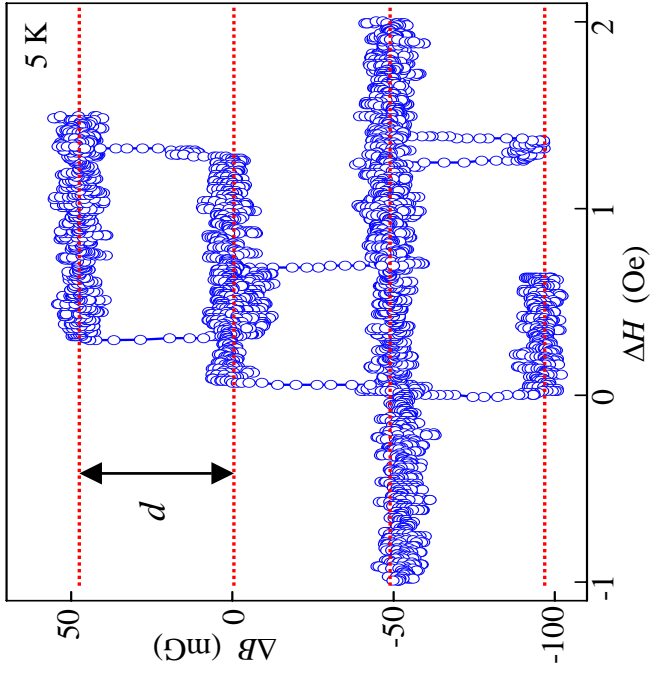

**Figure 3. Jumps of a domain wall between equivalent lattice sites.** The data were taken during a very slow sweep ($\approx 1$ h) required to achieve the sub-atomic resolution. For such time intervals, relaxation processes lead to irreproducible changes in the domain structure usually far away from the detection site (this can be seen as occasional DW jumps at a constant $H$). On the graph, this results in the same position of a DW for different values of $H$. For clarity, we subtracted a small smooth background in $B$ associated with changes in the local stray field induced by other domains.